\documentclass[twoside]{article}
\usepackage{fleqn,espcrc2}

\newcommand {\be}{\begin{equation}}
\newcommand {\ee}{\end{equation}}
\newcommand {\beq}{\begin{eqnarray}}
\newcommand {\eeq}{\end{eqnarray}}
\newcommand {\pr}{\partial}
\newcommand {\lf}{\lefteqn}
\newcommand {\sg}{\sigma}
\newcommand {\sgm}{\sigma^{'}}
\newcommand {\al}{\alpha}

\newcommand {\dl}{\delta}
\newcommand {\ep}{\epsilon}
\newcommand {\non}{\nonumber}

\newcommand{\AmS}{{\protect\the\textfont2
  A\kern-.1667em\lower.5ex\hbox{M}\kern-.125emS}}

\title{Bihamiltonian approach to the closed string model in the background
fields}

\author{V. D. Gershun\address{Institute of Theoretical
Physics, NSC Kharkov Institute of Physics and Technology, \\ P.O. Box
310108, 1 Akademicheskaya St., Kharkov,
Ukraine}\thanks{gershun@kipt.kharkov.ua}}
\begin{document} \begin{abstract}
 The closed string model in the background gravity field and the
 antisymmetric B-field is considered as the bihamiltonian system in
assumption,that string model is the integrable model for particular kind
of the background fields. It is shown, that bihamiltonity is origin of two
types of the T-duality of the closed string models. The dual nonlocal
Poisson brackets, depending of the background fields and of their
derivatives, are obtained.  The integrability condition is formulated as
the compatibility of the bihamoltonity condition and the Jacobi identity
of the dual Poisson bracket. It is shown, that the dual brackets and dual
hamiltonians can be obtained from the canonical (PB) and from the initial
hamiltonian by imposing of the second kind constraints on the initial
dynamical system, on the closed string model in the constant background
fields, as example.  The closed string model in the constant background
fields is considered without constraints, with the second kind constraints
and with first kind constraints as the B-chiral string. The two particles
discrete closed string model is considered as two relativistic particle
system to show the difference between the Gupta-Bleuler method of the
quantization with the first kind constraints and the quantization of the
Dirac bracket with the second kind constraints.

\vspace{1pc}
\end{abstract}

\maketitle
\section{Introduction}
  The bihamiltonian approach \cite{1,2,3} to the integrable systems
was initiated by Magri \cite{4} for the investigation of the
integrability of the KdV equation. A finite dimensional dynamical system
with 2N degrees of freedom $x^{a}$, $a=1...2N$ is integrable, if it is
described by the set of the $n$ integrals of motion $F_{1},..F_{n}$ in
involution under some Poisson bracket (PB) \be{\{F_{i}, F_{k}\}_{PB}=0}\ee
The dynamical system is completely solvable, if $n=N$. Any of the integral
of motion (or any linear combination of them) can be considered as the
hamiltonian \be{H_{k}:= F_{k}}\ee The bihamiltonity condition has
following form \be\label{3}\lf{\dot x^{a}= \frac{dx^{a}}{dt}=
\{x^{a},H_{1}\}_{1}=...= \{x^{a},H_{N}\}_{N}}\ee The hierarchy of new (PB)
is arose in this connection.  \be{\{ ,\}_{1}, \{ ,\}_{2},..\{ ,\}_{N}}\ee
The hierarchy of new dynamical systems is arose under the new time
coordinates $t_{k}$.
\be\label{5}\lf{\frac{dx^{a}}{dt_{n+k}}=\{x^{a},H_{n}\}_{k+1}}\ee
 The new equations of motion are describe the new dynamical systems,
which are dual to the original system, with the dual set of the integrals
of motion. The dual set of the integrals of motion can be obtained from
the original it by the mirror transformations and by the contraction of
the integrals of motion algebra. The contraction of the integral of motion
algebra means, that the dynamical system is belong to the orbits of
corresponding generators and is describe the invariant subspace. The set
of the commuting integrals of motion is belong to Cartan subalgebra of this
algebra. Consequently, duality is property of the integrable
models. KdV equation is one of the most interesting examples of the
infinite dimensional integrable mechanical systems with soliton solutions.
We are considered the dynamical systems with constraints. In this case,
first kind constraints are the generators of gauge the transformations and
they are integrals of motion.  First kind constrains $F_{k}(x^{a})\approx
0$, $k=1, 2...$ form the algebra of constraints under some (PB).
\be\{F_{i}, F_{k}\}_{PB}= C_{ik}^{l}F_{l}\approx 0\ee The structure
functions $C_{ik}^{l}$ may be functions of the phase space coordinates in
general case.  The second kind constraints $f_{k}(x^{a}) \approx 0$ are
the representations of the first kind constraints algebra.  The second
kind constraints is defined by the condition $\{f_{i}, f_{k}\}=C_{ik} \ne
0$.  The reversible matrix $C_{ik}$ is not constraint and it is function
of phase space coordinates also. The second kind constraints take part in
deformation of the $\{ ,\}_{PB}$ to the Dirac bracket $\{ , \}_{D}$. As
rule, such deformation leads to the nonlinear and to the nonlocal
brackets. First kind constraints are imposed upon the vector states under
the quantization: $F_{k}|\Psi> =0$.  The same spectrum of the excitations
and of the wave functions are obtained under the Gupta- Bleuer method of
the quantization.  One-half of the second kind constraints can be
considered as first kind constraints and they must be imposed upon vector
states in Gupta- Bleuer method of quantization. The (PB) is not deformed
to the Dirac bracket in this connection.  The bihamiltonity condition
leads to the dual (PB), which are nonlinear and nonlocal brackets as rule.
We suppose, that the dual brackets can be obtained from the initial
canonical bracket under the imposition of the second kind constraints. We
make this conclusion from the consideration of two dynamical models,
closed string model in the constant background fields and two particles
discrete closed string model, as examples.  The Gupta- Bleuer method of
the quantization may be more preferable in some case, if the dual bracket
is nonlocal. We have applied $\cite{5}$ bihamiltonity approach to the
investigation of the integrability of the closed string model in the
arbitrary background gravity field and antisymmetric B-field.  The
bihamiltonity condition and the Jacobi identities for the dual brackets
have considered as the integrability condition for a closed string model.
They led to some restrictions on the background fields. The local dual
(PB) of the similar type have considered $\cite{6}$ in the application to
the hamiltonian hydrodynamical models.  The (PB) of the hydrodynamical
type for the phase coordinate functions $u^{i}(x,t)$ is defined by formula
\beq\ \{u^{i}(x),u^{k}(y)\}=g^{ik}(u(x)){\pr_x}\delta (x-y) \\
+b^{ik}_{l}(u(x)){u^{l}_x}\delta (x-y)\non\eeq There $g^{ik}(u),
b^{ik}_{l}(u)$ are the arbitrary functions of the phase space coordinates
and $u_{x}={\pr_x}{u}$. The Jacobi identity is satisfied under the
following conditions:\\ 1. Tensor $g^{ik}$ is symmetric tensor and it is
define some metric on the phase space. \\ 2. $
b^{ik}_{l}(u)=-g^{ij}\Gamma ^{k}_{jl}(u)$ and connection $\Gamma^{k}_{jl}$
is consistent to metric $g^{ik}$ and it has zero curvature and zero
torsion. \\ Therefore, there are such local coordinates, that $g^{ik}$=
const, $b^{k}_{jl}=0$. This (PB) was used for description of the
hamiltonian system of the hydrodynamical type. That is systems with
functionals of the hydrodynamical type. The density of this functionals
does not depend of the derivatives $u^{k}_{x}, u^{k}_{xx},..$ and
hamiltonian is functional of the hydrodynamical type also.\\ In opposite
this models, the functionals of the closed string model is depended of the
derivatives of the string coordinates. As result, we need to introduce
additional nonlocal term with the step function $\ep (x-y)=2{\pr ^{-1}_{x}
\delta (x-y)}$, which is the origin of the difficulty of the Jacobi
identity proof. \\  The plan of the paper is following. In the second
section we are considered closed string model in the arbitrary background
gravity field and antisymmetric B-field as the bihamiltonian system. We
suppose, that this model is integrable model for some configurations of
the background fields. The bihamiltonity condition and the Jacobi
identities for the dual (PB) must be result to the integrability
condition, which is restrict the possible configurations of the background
fields. The well known examples of the integrable gravity models with the
gravity metric tensor, which is depended of one or of two variables only.
In this paper we are assumed the metric dependence of the arbitrary number
of the variables for generality and we did not analyzed the particular
cases of the metric dependence. In the third section we are considered
three examples of the closed string model in the constant background
fields: without constraints, with the second kind constraints and the
B-chiral string with the first kind constraints. In the four section we
are considered two particles discrete closed string model to show the
difference between the (PB) structure under the Gupta-Bleuer method of the
quantization and under the quantization of the system with the second kind
constraints.
\section {Closed string
in the background fields} The closed string in the background gravity
field and the antisymmetric B-field is described by first kind constraints
\beq\label{8}\lf{h_{1}=\frac{1}{2}g^{ab}(x)[p_{a}-\al
B_{ac}(x)x^{'c}][p_{b}-\al B_{bd}(x)x^{'d}]}\non \\
\lf{+\frac{1}{2}g_{ab}(x)x^{'a}x^{'b}\approx 0,\,\,
h_{2}=p_{a}x^{'a}\approx 0}\eeq where $a,b
=0,1,...D-1$, $x^{a}(\sg), p_{a}(\sg)$ are the periodical functions on
$\sg$ with the period on $\pi$ , $\al$ -arbitrary parameter. The
original (PB) are the symplectic (PB) \beq\label{9}\lf{\{x^{a}(\sg
),p_{b}(\sgm )\}_{1}=\dl _{b}^{a}\dl (\sg -\sgm )} \\
\lf{\{x^{a},x^{b}\}_{1}=\{p_{a},p_{b}\}_{1}=0}\non\eeq The hamiltonian
equations of motion of the closed string, in the arbitrary background
gravity field and antisymmetric B-field under the hamiltonian
$H_{1}=\int\limits_{0}^{\pi}h_{1}d\sg$
and (PB) $\{,\}_{1}$, are
\beq\label{10}\lf{\dot x^{a}= g^{ab}[p_{b}-\al
B_{bc}x^{'c}]}\\
\lf{\dot p_{a}= \al B_{ab}g^{bc}p_{c}^{'}+[g_{ab}-\al^{2}
B_{ac}g^{cd}B_{db}]x^{''b}-}\non \\
\lf{-\frac{1}{2}\frac{\pr g^{bc}}{\pr x^{a}}p_{b}p_{c}-\al\frac{\pr}{\pr
x^{a}}(B_{bd}g^{dc})x^{'b}p_{c}+}\non \\
\lf{+\al\frac{\pr}{\pr x^{b}}(B_{ad}g^{dc})x^{'b}p_{c}-}\non \\
\lf{-\frac{1}{2}\frac{\pr}{\pr
x^{a}}[g_{bc}-\al^{2}B_{bd}g^{de}B_{eb}]x^{'b}x^{'c}+}\non \\
\lf{+\frac{\pr}{\pr
x^{b}}[g_{ac}-\al^{2}B_{ad}g^{de}B_{ec}]x^{'b}x^{'c}}\non\eeq
The dual (PB) are obtained from the bihamiltonity condition
\beq\label{11}\lf{\dot x^{a}=\{x^{a},\int\limits_{0}^{\pi}h_{1}d\sgm\}_{1}=
\{x^{a},\int\limits_{0}^{\pi}h_{2}d\sgm\}_{2}}\\
\lf{\dot p_{a}=\{p_{a},\int\limits_{0}^{\pi}h_{1}d\sgm\}_{1}=
\{p_{a},\int\limits_{0}^{\pi}h_{2}d\sgm\}_{2}}\non\eeq
and they have following form
\beq\label{12}\lf{\{A(\sg ),B(\sgm )\}_{2}=}\\
\lf{\frac{\pr A}{\pr x^{a}}\frac{\pr B}{\pr x^{b}}[[\omega^{ab}(\sg
)+\omega^{ab}(\sgm )]\ep (\sgm -\sg )+}\non \\ \lf{[\Phi^{ab}(\sg
)+\Phi^{ab}(\sgm )]\frac{\pr}{\pr \sgm }\dl (\sgm -\sg )+}\non \\
\lf{[\Omega^{ab}(\sg )+\Omega^{ab}(\sgm )]\dl (\sgm -\sg )]+}\non\eeq
\beq\lf{\frac{\pr A}{\pr p_{a}}\frac{\pr B}{\pr p_{b}}[[\omega_{ab}(\sg
)+\omega_{ab}(\sgm )]\ep (\sgm -\sg )+}\non \\
\lf{[\Phi_{ab}(\sg )+\Phi_{ab}(\sgm )]\frac{\pr}{\pr \sgm }\dl (\sgm -\sg
)+}\non \\ \lf{[\Omega_{ab}(\sg )+\Omega_{ab}(\sgm )]\dl (\sgm -\sg
)]+}\non\eeq  \beq\lf{[\frac{\pr A}{\pr x^{a}}\frac{\pr B}{\pr
p_{b}}+\frac{\pr A}{\pr p_{b}}\frac{\pr B}{\pr x^{a}}][[\omega_{b}^{a}(\sg
)+\omega_{b}^{a}(\sgm )]\ep (\sgm -\sg )}\non \\ \lf{+[\Phi_{b}^{a}(\sg
)+\Phi_{b}^{a}(\sgm )]\frac{\pr}{\pr \sgm }\dl (\sgm -\sg )]+}\non \\
\lf{[\frac{\pr A}{\pr x^{a}}\frac{\pr B}{\pr p_{b}}-\frac{\pr
A}{\pr p_{b}}\frac{\pr B}{\pr x^{a}}][\Omega _{b}^{a}(\sg )+
\Omega_{b}^{a}(\sgm )]\dl (\sgm -\sg )}\non\eeq The arbitrary functions
$A, B, \omega, \Phi, \Omega$ are the functions of the $x^{a}(\sg ),
p_{a}(\sg )$.  The functions $\omega^{ab}, \omega_{ab}$,
$\Phi^{ab},\Phi_{ab}$ are the symmetric functions on $a, b$ and
$\Omega^{ab}, \Omega_{ab}$ are the antisymmetric functions to satisfy the
condition $\{A, B\}_{2}=-\{B,A\}_{2}$.  The equations of motion under the
hamiltonian $H_{2}=\int\limits_{0}^{\pi} h_{2}(\sgm)d\sgm$ and (PB)
$\{, \}_{2}$ are \beq\label{13}\lf{\dot
x^{a}=-2\omega_{b}^{a}x^{b}+4\omega^{ab}p_{b}+2\Phi^{ab}p_{b}^{''}-}\\
\lf{-2\Phi_{b}^{a}x^{''b}+2\Omega_{b}^{a}x^{'b}-2\Omega^{ab}p_{b}^{'}+}
\non \\ \lf{\int\limits_{0}^{\pi}d\sgm [\omega_{b}^{a}x^{'a}+\frac{\pr
\omega^{ac}}{\pr x^{b}}x^{'b}p_{c}+\frac{\pr \omega^{ac}}{\pr
p_{b}}p_{b}^{'}p_{c}]\ep (\sgm -\sg )}\non \\
\lf{+(\frac{\pr \Phi^{ac}}{\pr x^{b}}x^{'b}+\frac{\pr \Phi^{ac}}{\pr p_{b}}
p_{b}^{'})p_{c}^{'}-}\non \\
\lf{-(\frac{\pr \Phi_{c}^{a}}{\pr x^{b}}x^{'b}+\frac{\pr \Phi_{c}^{a}}{\pr
p_{b}} p_{b}^{'})x^{'c}}\non\eeq
\beq\label{14}\lf{\dot
p_{a}=-2\omega_{ab}x^{b}-2\Phi_{ab}x^{''b}+2\Omega_{ab}x^{'b}+}\\
\lf{+4\omega_{a}^{b}p_{b}+2\Phi_{a}^{b}p_{b}^{''}+2\Omega_{a}^{b}p_{b}^{'}+
\int\limits_{0}^{\pi}d\sgm
(\omega_{ab}x^{'b}-}\non \\ \lf{-\frac{\pr^{2}\omega^{cd}}{\pr x^{a}\pr
x^{b}}x^{'b}p_{c}p_{d}-\frac{\pr\omega^{bc}}{\pr x^{a}}p_{b}p_{c}^{'})\ep
(\sgm -\sg )}\non\eeq
\beq\lf{-(\frac{\pr \Phi_{ac}}{\pr x^{b}}x^{'b}+\frac{\pr \Phi_{a}^{c}}{\pr
p_{b}}p_{b}^{'})x^{'c}+}\non \\
\lf{+(\frac{\pr \Phi_{a}^{c}}{\pr x^{b}}x^{'b}+\frac{\pr \Phi_{a}^{c}}
{\pr p_{b}}p_{b}^{'})p_{c}^{'}}\non\eeq
The bihamiltonity condition (\ref{11}) is led to the two constraints
\beq\label{15}\lf{
-2\omega_{b}^{a}x^{b}+4\omega^{ab}p_{b}+2\Phi^{ab}p_{b}^{''}-}\\
\lf{-2\Phi_{b}^{a}x^{''b}+2\Omega_{b}^{a}x^{'b}-2\Omega^{ab}p_{b}^{'}+}
\non \\ \lf{\int\limits_{0}^{\pi}d\sgm [\omega_{b}^{a}x^{'a}+\frac{\pr
\omega^{ac}}{\pr x^{b}}x^{'b}p_{c}+\frac{\pr \omega^{ac}}{\pr
p_{b}}p_{b}^{'}p_{c}]\ep (\sgm -\sg )}\non \\
\lf{+(\frac{\pr \Phi^{ac}}{\pr x^{b}}x^{'b}+\frac{\pr \Phi^{ac}}{\pr p_{b}}
p_{b}^{'})p_{c}^{'}-}\non \\
\lf{-(\frac{\pr \Phi_{c}^{a}}{\pr x^{b}}x^{'b}+\frac{\pr \Phi_{c}^{a}}{\pr
p_{b}} p_{b}^{'})x^{'c}=g^{ab}p_{b}-\al g^{ab}B_{bc}x^{'c}}\non\eeq
\beq\lf{
-2\omega_{ab}x^{b}-2\Phi_{ab}x^{''b}+2\Omega_{ab}x^{'b}+}\\
\lf{+4\omega_{a}^{b}p_{b}+2\Phi_{a}^{b}p_{b}^{''}+2\Omega_{a}^{b}p_{b}^{'}+
\int\limits_{0}^{\pi}d\sgm
(\omega_{ab}x^{'b}-}\non \\ \lf{-\frac{\pr^{2}\omega^{cd}}{\pr x^{a}\pr
x^{b}}x^{'b}p_{c}p_{d}-\frac{\pr\omega^{bc}}{\pr x^{a}}p_{b}p_{c}^{'})\ep
(\sgm -\sg )}\non\eeq
\beq\label{16}\lf{-(\frac{\pr \Phi_{ac}}{\pr x^{b}}x^{'b}+\frac{\pr
\Phi_{a}^{c}}{\pr p_{b}}p_{b}^{'})x^{'c}+}\non \\ \lf{+(\frac{\pr
\Phi_{a}^{c}}{\pr x^{b}}x^{'b}+\frac{\pr \Phi_{a}^{c}} {\pr
p_{b}}p_{b}^{'})p_{c}^{'}=}\non\eeq \beq\lf{=\al
B_{ab}g^{bc}p_{c}^{'}+[g_{ab}-\al^{2}B_{ad}g^{dc}B_{cb}]x^{'b}-}\non \\
\lf{-\frac{1}{2}\frac{\pr g^{bc}}{\pr x^{a}}p_{b}p_{c}-\al\frac{\pr}{\pr
x^{a}}(B_{bd}g^{dc})x^{'b}p_{c}+}\non \\
\lf{+\al \frac{\pr}{\pr x^{b}}(B_{ad}g^{dc})x^{'b}p_{c}-}\non\eeq
\beq\lf{-\frac{1}{2}\frac{\pr}{\pr
x^{a}}[g_{bc}-\al^{2}B_{bd}g^{de}B_{eb}]x^{'b}x^{'c}+}\non \\
\lf{\frac{\pr}{\pr
x^{b}}[g_{ac}-\al^{2}B_{ad}g^{de}B_{ec}]x^{'b}x^{'c}}\non\eeq
In really, there is the list of the constraints depending on the possible
choice of the unknown functions $\omega, \Omega, \Phi$. In the general
case, there are as the first kind constraints as the second kind
constraints too. Also, it is possible to solve the constraints equations
as the equations for the definition of the functions $\omega, \Omega,
\Phi$.  We are considered last possibility and we obtained the following
consistent solution of the bihamiltonity condition.
\beq\label{17}\lf{\Phi^{ab}=0,\Omega^{ab}=0,\Phi_{b}^{a}=0,
\omega^{ab}=Cg^{ab}}\\
\lf{\omega_{ab}=\frac{C}{2}\frac{\pr^{2} g^{cd}}{\pr x^{a}\pr
x^{b}}p_{c}p_{d},\,\,\omega_{b}^{a}=-C\frac{\pr g^{ac}}{\pr
x^{b}}p_{c},}\non \\
\lf{\Phi_{ab}=-\frac{1}{2}[g_{ab}-\al^{2}B_{ac}g^{cd}B_{db}],}\non \\
\lf{\Omega_{ab}=\frac{1}{2}(\frac{\pr \Phi_{bc}}{\pr x^{a}}-\frac{\pr
\Phi_{ac}}{\pr x^{b}})x^{'c}-}\non \\
\lf{-(\frac{\pr \Omega_{b}^{c}}{\pr x^{a}}-\frac{\pr \Omega_{a}^{c}}{\pr
x^{b}})p_{c}, \frac{\pr \omega^{ab}}{\pr p_{c}}=0},\non \\
\lf{2\Omega_{b}^{a}=-\al g^{ac}B_{ca}, \frac{\pr g^{ab}}{\pr
x^{c}}x^{c}=ng^{ab}, C=\frac{1}{2(n+2)}}\non\eeq
The metric tensor $g^{ab}(x)$ is the homogeneous function of $x^{a}$ order
$n$ and $C$ is arbitrary constant.  In the difference of the (PB) of the
hydrodynamical type, we are needed to introduce the separate (PB) for the
coordinates of the Minkowski space and for the momenta because, the
gravity field is not depend of the momenta.  Although, this difference is
vanished under the such constraint as $f(x^{a},p_{a})\approx 0$. One can
see, that the main term in the (PB) with the metric tensor is the term
with the step function $\epsilon (\sg- \sgm)$. The functions
$\omega_{a}^{b}$, $\Omega{ab}$ are proportional to the connection and to
the torsion. The function $\omega_{ab}$ is proportional to the curvature
and the product of the connections.
 The dual (PB) for the phase
space coordinates are \beq\lf{\{x^{a}(\sg ),x^{b}(\sgm )\}_{2}=
[\omega^{ab}(\sg )+\omega^{ab}(\sgm )]\ep (\sgm -\sg )}\non\eeq
\beq\label{18}\lf{\{p_{a}(\sg ),p_{b}(\sgm )\}_{2}= [\omega_{ab}(\sg
)+\omega_{ab}(\sgm )]\ep (\sgm -\sg )+}\non \\
\lf{[\Phi_{ab}(\sg )+\Phi_{ab}(\sgm )]\frac{\pr}{\pr \sgm }\dl (\sgm -\sg
)+}\non \\ \lf{[\Omega_{ab}(\sg )+\Omega_{ab}(\sgm )]\dl (\sgm
-\sg)}\eeq \beq\lf{\{x^{a}(\sg ),p_{b}(\sgm
)\}_{2}= [\omega_{b}^{a}(\sg )+\omega_{b}^{a}(\sgm )]\ep (\sgm -\sg )}\non
\\ \lf{+[\Omega _{b}^{a}(\sg )+ \Omega_{b}^{a}(\sgm )]\dl (\sgm -\sg
)}\non\eeq \beq\lf{\{p_{a}(\sg ),x^{b}(\sgm )\}_{2}=[\omega_{b}^{a}(\sg
)+\omega_{b}^{a}(\sgm )]\ep (\sgm -\sg )}\non \\ \lf{+[\Omega _{b}^{a}(\sg
) + \Omega_{b}^{a}(\sgm )]\dl (\sgm -\sg )}\non\eeq The functions
$\omega^{ab}(x),\omega_{ab}(x)$,$\Phi_{ab}(x),\Omega_{ab}(x)$,
$\omega^{a}_{b}(x),\Omega^{a}_{b}(x)$ is defined in (17).
 It is rather easy to prove the Jacobi identities for the local part of
the dual (PB) $\{, \}_{2}$. It does not understand, how to prove the Jacobi
identities for the nonlocal part of it.  The principal term of the Jacobi
identities with the step functions only is the term with the structure
function $\omega^{ab}(x)$.
\beq\lf{[\frac{\pr g^{ab}(\sg )}{\pr x^{d}}[g^{dc}(\sg
 )+g^{dc}(\sg^{''} )]-}\\ \lf{-\frac{\pr g^{ac}(\sg )}{\pr
x^{d}}[g^{db}(\sg )+g^{db}(\sgm )]]\ep (\sgm -\sg )\ep (\sigma^{''}-\sg
)+}\non\eeq \beq\lf{+[\frac{\pr g^{cb}(\sg )}{\pr x^{d}}[g^{da}(\sgm
)+g^{da}(\sg )]-}\non \\ \lf{-\frac{\pr g^{ab}(\sg )}{\pr
x^{d}}[g^{dc}(\sgm )+g^{dc}(\sg^{''} )]]\ep (\sg -\sgm )\ep
(\sigma^{''}-\sgm )+}\non\eeq \beq\lf{[\frac{\pr g^{ac}(\sg^{''} )}{\pr
x^{d}}[g^{db}(\sg^{''} )+g^{db}(\sgm )]-\frac{\pr
g^{cb}(\sg^{''} )}{\pr x^{d}}[g^{da}(\sg^{''} )}\non \\ \lf{+g^{da}(\sg
)]]\ep (\sg -\sg^{''})\ep (\sgm -\sg^{''})=0}\non\eeq It is possible to
reduce this condition to the unique equation \beq\lf{[\frac{\pr g^{ab}(\sg
)}{\pr x^{d}}[g^{dc}(\sg )+g^{dc}(\sg^{''} )]-\frac{\pr g^{ac}(\sg )}{\pr
x^{d}}[g^{db}(\sg )}\non\\ \lf{+g^{db}(\sgm )]]\ep (\sgm -\sg )\ep
(\sg^{''}-\sg )=0}\eeq One of the possible way of the solution of
this problem is the consideration of the metric tensor on the phase space
to count the contribution of the structure functions $\Omega$ and $\Phi$
to this expression. The second possible way is the consideration of the
second kind constraints from the list of the constraints
 (\ref{15}),(\ref{16}), instead of the solution (\ref{17}).  It is
necessary to introduce this constraints to the initial model with the
hamiltonian $H_{1}$, the (PB) $\{, \}_{1}$ and to obtain the bihamiltonity
condition in this case. As we will see later on the some examples, the
second kind constraints of the $f(x^{a},p_{a})\approx 0$ type can lead to
the nonlocal Dirac bracket with the step function from the one side, and
they can introduce the dependence of the metric tensor of the momenta on
the solutions of this constraints from the other side.  At present, this
problems are under the consideration.  \section{Constant background
fields} The bihamiltonity condition (\ref{15}),(\ref{16}) is reduced to
the following constraints on the phase space \beq\label{21}\lf{
-2\omega_{b}^{a}x^{b}+4\omega^{ab}p_{b}+2\Phi^{ab}p_{b}^{''}-
2\Phi_{b}^{a}x^{''b}+}\non \\
\lf{+2\Omega_{b}^{a}x^{'b}=g^{ab}p_{b}-\al
g^{ab}B_{bc}x^{'c},\,\Omega^{ab}=0}\\
\lf{-4\omega_{ab}x^{b}-2\Phi_{ab}x^{''b}+2\Omega_{ab}x^{'b}+
4\omega_{a}^{b}p_{b}+}\non \\
\lf{+2\Phi_{a}^{b}p_{b}^{''}+2\Omega_{a}^{b}p_{b}^{'}=\al
B_{ab}g^{bc}p_{c}^{'}+}\non \\
\lf{+[g_{ab}-\al ^{2}B_{ac}g^{cd}B_{db}]x^{''b}}\non\eeq
There is the unique solution without constraints
\beq\label{22}\lf{\omega^{ab}=\frac{1}{4}g^{ab},2\Omega_{b}^{a}=-\al
g^{ac}B_{cb}=\al B_{bc}g^{ca},}\non \\
\lf{-2\Phi_{ab}=g_{ab}-\al^{2}B_{ac}g^{cd}B_{db}}\eeq
The rest structure functions are equal zero.
In this section we are supplemented the bihamiltonity condition (\ref{11})
by the mirror transformations of the integrals of motion.  \beq\lf{\dot
x^{a}=\{x^{a},\int\limits_{0}^{\pi}h_{1}d\sgm\}_{1}=
\{x^{a},\int\limits_{0}^{\pi}{\pm h_{2}}d\sgm\}_{\pm 2}}\eeq
 The dual (PB) are
\beq\label{24}\lf{\{x^{a}(\sg ),x^{b}(\sgm
)\}_{\pm 2}=\pm\frac{1}{2}g^{ab}\ep (\sgm -\sg )}\\ \lf{\{x^{a}(\sg
),p_{b}(\sgm )\}_{\pm 2}=\mp\al g^{ac}B_{cb}\dl (\sgm -\sg )}\non\\
\lf{\{p_{a}(\sg ),p_{b}(\sgm )\}_{\pm 2}=}\non \\
\lf{=\mp[g_{ab}-\al ^{2}B_{ac}g^{cd}B_{db}]\frac{\pr}{\pr \sgm }\dl (\sgm
-\sg )}\non\eeq The dual dynamical system \be\label{25}\lf{\dot
x^{a}=\{x^{a}, \pm H_{2}\}_{1}=\{x^{a}, H_{1}\}_{\pm 2}}\ee is the
left(right) chiral string \be\label{26}\lf{\dot x^{a}=\pm x^{'a},\,\, \dot
p_{a}=\pm p_{a}^{'}}\ee In the terms of the Virasoro operators \beq\lf{
L_{k}=\frac{1}{4\pi}\int\limits_{0}^{\pi}(h_{1}+h_{2})e^{ik\sg}d\sg} \\
\lf{\bar
L_{k}=\frac{1}{4\pi}\int\limits_{0}^{\pi}(h_{1}-h_{2})e^{ik\sg}d\sg}\non\eeq
the first kind constraints form the $Vir\oplus Vir$ algebra under the (PB)
$\{, \}_{1}$.
\beq\label{28}\lf{\{L_{n},L_{m}\}_{1}=-i(n-m)L_{n+m}}\non \\
\lf{\{\bar L_{n},\bar L_{m}\}_{1}=-i(n-m)\bar L_{n+m}}\label{com}\non \\
\lf{\{L_{n},\bar L_{m}\}_{1}=0}\eeq
The dual set of the integrals of motion is obtained from initial it by the
mirror transformations
\beq\lf{
H_{1}\to \pm H_{2}, L_{0}\to \pm L_{o}, \bar L_{0}\to \mp \bar L_{0}, \tau
\to \sg} \eeq
and by the contraction of the first kind constraints algebra $L_{n}=0$, or
$\bar L_{n}=0$, $n\ne 0$.
\subsection{ Second kind constraints}
Another way to obtain the dual brackets is the imposition of the second
kind constraints on the initial dynamical system, by such manner, that
$F_{i}=F_{k}$ for $i\ne k, i, k = 1,2,...$ on the constraints surface
$f(x^{a},p_{a})=0$. Let us consider the closed string model with
$B_{ab}=0$ for simplicity.
\beq\label{30}\lf{h_{1}=\frac{1}{2}g^{ab}p_{a}p_{b}+
\frac{1}{2}g_{ab}x^{'a}x^{'b}},\non \\ \lf{h_{2}=p_{a}x^{'a}}\eeq
The constraints $f^{(-)}_{a}(x,p)=p_{a}-g_{ab}x'^{b}\approx 0$ or
$f^{(+)}_{a}=p_{a}+g_{ab}x'^{b}\approx 0$(do not simultaneously) are the
second kind constraints.  \beq\label{31}\lf{\{f^{(\pm)}_{a}(\sg),
f^{(\pm)}_{b}(\sgm)\}_{1}=C^{(\pm)}_{ab}(\sg -\sgm)=}\non\\
\lf{=\pm 2g_{ab}\frac{\pr}{\sgm}\delta (\sgm -\sg)}\eeq The inverse matrix
$(C^{(\pm)})^{-1}$ has following form
\beq\label{32}\lf{C^{(\pm)ab}(\sg
-\sgm)=\pm\frac{1}{4}g^{ab}\epsilon(\sgm -\sg)}\eeq There is only one set
of the constraints, because consistency condition \beq\lf{\{f^{(\pm)}(\sg),
H_{1}\}_{1}= f^{'(\pm)}(\sg)\approx 0, ...}\non\\
\lf{\{f^{(\pm)(n)}(\sg), H_{1}\}_{1}= f^{(\pm)(n+1)}(\sg)
\approx 0}\eeq is not produce the new sets of constraints.  By using the
standard definition of the Dirac bracket, we are obtained following Dirac
brackets for the phase space coordinates.  \beq\label{34}\lf{\{x^{a}(\sg),
x^{b}(\sgm)\}_{D}=\pm\frac{1}{4}g^{ab}\epsilon(\sgm -\sg),} \\
\lf{\{p_{a}(\sg),
p_{b}(\sgm)\}_{D}=\mp\frac{1}{2}g_{ab}\frac{\pr}{\sgm}\delta(\sgm
-\sg),}\non\\ \lf{\{x^{a}(\sg),
p_{b}(\sgm)\}_{D}=\frac{1}{2}\delta^{a}_{b}\delta(\sgm -\sg)}\non\eeq The
equations of motion under the hamiltonians $H_{1}=h_{1},H_{2}= h_{2}$ and
Dirac bracket \beq\label{35}\lf{\dot x^{a}=\{x^{a}, H_{1}\}_{D}=\{x^{a},
H_{2}\}_{D}=g^{ab}p_{b}=\pm x'^{a}}\\ \lf{\dot p_{a}=\{p_{a},
H_{1}\}_{1}=\{p_{a}, H_{2}\}_{D}=g_{ab}x'^{b}=\pm p'_{a}}\non\eeq are
coincide on the constraints surface. The dual brackets $\{, \}_{\pm 2}$
are coincide with the Dirac brackets also. The contraction of the algebra
of the first kind constraints means that the integrals of motion
$H_{1}=H_{2}$ are coincide on the constraints surface too.
\subsection{B-chiral string} Let us consider the following constraint from
the list (\ref{24}) \be\label{36}\lf{\varphi_{a}=p_{a}+\beta
B_{ab}x^{'b}\approx 0}\ee The consistency condition \beq\lf{\{\varphi_{a},
H_{1}\}_{1}=(\al +\beta )B_{ab}g^{bc} \varphi_{c}^{'}+}\non \\
\lf{+[g_{ab}-(\al +\beta )^{2}B_{ac}g^{cd}B_{db}]x^{''b}\approx 0}\eeq
show, that under the additional condition on the B-field
\be\label{38}\lf{g_{ab}=(\al +\beta )^{2}B_{ac}g^{cd}B_{db}},\ee the
constraints $\varphi_{a}\approx 0$ are first kind constraints.  The motion
equations are \beq\lf{\dot x^{a}=-(\al +\beta )g^{ab}B_{bc}x^{'c}}\\
\lf{\dot p_{a}=-(\al +\beta )B_{ab}g^{bc}p_{c}^{'},\,\,\ddot
x^{a}=x^{''a}}\non\eeq This model is the bihamiltonian model under (PB)
(\ref{26}) also. The B-chiral string model is dual to the chiral model
also.  \section{Two particles discrete string} In this section we are
considered two particles discrete closed string model, as two relativistic
particles model, to show the difference between the Dirac brackets
quantization and the Gupta-Bleuer quantization methods.  Two and three
pieces discrete string in Gupta-Bleuer method of quantization was
considered in the paper \cite{7} and it is described by the following
constraints
\beq\label{40}\lf{h=\frac{1}{4}(p^{2}+q^{2})+\omega_{0}^{2}r^{2}\approx
0,}\non\\ \lf{f_{1}=pq\approx 0, f_{3}=qr\approx 0,}\\ \lf{f_{2}=pr\approx
0, f_{4}=\frac{1}{4}q^{2}-\omega_{0}^{2}r^{2}}\non\eeq This model is the
two relativistic particles system with the oscillator interaction. The
constraints (\ref{40}) are two particles discrete analog of the Virasoro
constraints and $p, q, r$ are the collective variables
$p^{a}=p^{a}_{2}+p^{a}_{1}$,$q^{a}=p^{a}_{2}-p^{a}_{1}$,
$r^{a}=x^{a}_{2}-x^{a}_{1}$.
Under the hamiltonian $H=h$ and the canonical (PB)
$\{r^{a},q^{b}\}=2\eta^{ab}$, the constraints $f_{i}$ are the second
kind constraints.  \beq\label{41}\lf{\{f_{1}, f_{2}\}=2p^{2}\ne 0,}\\
\lf{\{f_{3}, f_{4}\}=q^{2}+4\omega_{0}^{2}r^{2}\ne 0}\non\eeq The string
coordinates are satisfied to following Dirac brackets.  \beq\lf{\{r_{a},
r_{b}\}_{D}=\frac{r_{b}q_{a}-r_{a}q_{b}}{q^{2}+\omega_{0}^{2}r^{2}},
\{q_{a}, q_{b}\}_{D}=4\omega_{0}\{r_{a}, r_{b}\},}\non\\
\lf{\{r_{a}, q_{b}\}_{d}=2(\eta_{ab}-\frac{p_{a}p_{b}}{p^{2}}-
-\frac{q_{a}q_{b}+4\omega_{0}^{2}r_{a}r_{b}}{q^{2}+ 4\omega_{0}^{2}r^{2}})}
\eeq
In the terms of the amplitudes $a_{a}^{(+)},a_{a}$ of the
equations of motion solutions
\beq\lf{r_{a}=a_{a}e^{i\omega_{0}\tau}
+a_{a}^{(+)}e^{-i\omega_{0}\tau},}\\
\lf{q_{a}=2i\omega_{0}(a_{a}e^{i\omega_{0}\tau}
-a_{a}^{(+)}e^{-i\omega_{0}\tau})}\non\eeq they have form \beq\lf{\{r_{a},
r_{b}\}_{D}=\frac{i(a_{b}^{(+)}a_{a}-a_{a}^{(+)}a_{b})}
{2\omega_{0}a_{k}^{(+)}a_{k}},}\non\\
\lf{\{q_{a}, q_{b}\}_{D}=2\omega_{0}\{r_{a}, r_{b}\},}\\
\lf{\{r_{a}, q_{b}\}_{D}=2(\eta_{ab}-\frac{p_{a}p_{b}}{p^{2}}-
\frac{a_{a}^{(+)}a_{b}+a_{b}^{(+)}a_{a}}{2a_{k}^{(+)}a_{k}})}\non\eeq
This Dirac brackets is possible to solve in the terms of the variables
$a,a^{(+)}$.
\beq\label{45}\lf{\{a_{a},
a_{b}^{(+)}\}_{D}=\frac{i}{2\omega_{0}}(\eta_{ab}-\frac{p_{a}p_{b}}
{p^{2}}-
\frac{ia_{a}^{(+)}a_{b}}{2\omega_{0}a_{k}^{(+)}a_{k}})}\eeq
The hamiltonian and the linear combinations of the constraints
$f_{1},.f_{4}$ have following form.
\beq\lf{H=\frac{1}{4}p^{2}+4\omega_{0}^{2}a_{k}^{(+)}a_{k},}\\
\lf{f_{1}=p_{k}a^{k}, f_{2}=p_{k}a_{k}^{(+)}, f_{3}=a_{k}a^{k},
f_{4}=a_{k}^{(+)}a^{(+)k}}\non\eeq
Under the quantization $[, ]\to i\{, \}_{D}$, $a_{k}, a_{k}^{(+)}\to$
operators $a_{k},a_{k}^{(+)}$, the commutation relation is
\beq\label{47}\lf{[a_{k},
a_{l}^{(+)}]=-\frac{1}{2\omega_{0}}(\eta_{kl}-\frac{p_{k}p_{l}}{p^{2}})+}\non
\\ \lf{+\frac{1}{2\omega_{0}}(a^{(+)i}a_{i})^{-1}a_{k}^{(+)}a_{l}}\eeq
Last term of the commutation relation has transposed Lorentz indeces
$k,l$. The wave function
\beq\lf{|\Psi_{n}(p)>=a_{k_{1}}^{(+)},
a_{k_{2}}^{(+)}...a_{k_{n}}^{(+)}\Psi_{k_{1}k_{2}...k_{n}}(p)|0>}\eeq of
the physical states must to be the own function of the hamiltonian $H$ on
the constraints surface. Let us consider the two
particles  excited state, for example.
\beq\lf{H|\Psi_{2}(p)>=
(\frac{1}{4}p^{2}-4\omega_{0}^{2})\Psi_{k_{1}k_{2}}(p)a_{k_{1}}^{(+)}
a_{k_{2}}^{(+)}|0>}\eeq
+ terms, which are proportional to the expressions
\beq\lf{p^{k_{1}}\Psi_{k_{1}k_{2}}(p),
\eta^{k_{1}k_{2}}\Psi_{k_{1}k_{2}}(p),
a_{l}^{(+)}a^{(+)l}\Psi_{k_{1}k_{2}}(p)}\non\eeq
Consequently, we must to impose additional conditions on the wave function
$p^{k}\Psi_{kl}=0$, $\eta^{kl}\Psi_{kl}=0$ to satisfy the request
about the own function. Last term $a_{i}^{(+)}a^{(+)i}\Psi_{kl}$ is
vanished on the constraints surface. In contrast to the nonlinear and to
the nonlocal Dirac bracket (\ref{45}), we have the canonical (PB) and two
first kind constraints $p_{k}a^{k}\approx 0$,$a_{k}a^{k}\approx 0$ in the
Gupta-Bleuer method of the quantization.  The first kind constraints $H,
f_{1}, f_{3}$ are imposed on the vector states and they are led to the
following equations on the wave function.
\beq\lf{(p^{2}-16\omega_{0}^{2})\Psi_{k_{1}k_{2}..k_{n}}(p)=0,}\\
\lf{p^{k}\Psi_{kk_{2}..k_{n}}(p)=0, \eta^{kl}\Psi_{kl..k_{n}}(p)=0}\non\eeq
\section{Relativistic particle in the constant electromagnetic field}
 The relativistic particle in the constant background electromagnetic field
is described by the Hamiltonian
\be\lf{H=\frac{1}{2}[(p_{a}+i\beta B_{ab}x_{b})^{2}+m^{2}]}\ee
The electromagnetic field is $A_{a}(x)=-2F_{ab}x_{b}$=$-B_{ab}x_{b}$. The
simplest constraint from (\ref{21}) is
\be\lf{\varphi_{a}=p_{a}+i\al B_{ab}x_{b}\approx 0}\ee
The consistency condition
\be\lf{\{\varphi_{a},H\}=i\al(\al +\beta )B_{ab}\varphi_{b}}\ee
shows that there is a unique set of constraints if $\al +\beta =
0$. They are the second class constraints
$\{\varphi_{a},\varphi_{b}\}=2i\al B_{ab}$. There is the
following algebra of the phase space coordinates under the Dirac
bracket
\beq\lf{\{x_{a},x_{b}\}_{D}=\frac{-i}{2\al}(B^{-1})_{ab},\{x_{a},p_{b}\}_
{D}=\frac{1}{2}\eta_{ab}}\non \\
\lf{\{p_{a},p_{b}\}_{D}=\frac{-i\al}{2}B_{ab}}\eeq
The motion equation under the Dirac bracket
\be\lf{\dot x_{a}+2i\al B_{ab}x_{b}=0,\, \dot p_{a}+2i\al
B_{ab}x_{b}=0}\ee has the solution $x_{a}(\tau )=\{e^{-2i\al
B\tau}\}_{ab}x_{b}(0)$. The quantization of the Dirac bracket
results in the following commutation relations.
\beq\label{56}\lf{[x_{a},x_{b}]=\frac{1}{2\al}(B^{-1})_{ab},\,[p_{a},p_{b}]=
\frac{\al}{2}B_{ab}}\non \\
\lf{[x_{a},p_{b}]=\frac{i}{2}\eta_{ab}}\eeq

\section{Acknowledgements} The author would like to thanks J. Lukierski
for the kind hospitality in the Wroclaw University and A.I. Pashnev for
the useful discussions.


\begin{thebibliography}{17}
\bibitem{1}     L. D. Fadeev and L. A. Takhtajan,
		 Hamiltonian methods in theory of solitons.
		 Springer, Berlin, (1987)
\bibitem{2}   J..A. Mitropolski, N. N. Bogolubov(Jr), A. K.
		 Prikarpatski, V. G. Samoylenko,
		 Integriruemie dinamicheskie
		 sistemi:spectralno-geometricheskie aspecti.
		 Naukova Dumka, Kiev, (1987)(in Russian).
\bibitem{3}     S. Okubo, A.  Das, Phys. Lett., B209, 311, (1988).
\bibitem{4}     F. A. Magri, Lect. Notes Phys., 19, 1156, (1978).
\bibitem{5}     V. D. Gershun, Nucl. Phys. B (Proc. Suppl.),102\&103, 71,
                (2001), hep-th/0103097.
		V. D. Gershun, Gravitation, cosmology and relativistic
		astrophysics, KNU, Kharkov, 69, (2001).
\bibitem{6}     B. A. Dubrovin, S. P. Novikov, Sov. Uspekhi Math., 44, 29,
                (1989).
\bibitem{7}     V. D. Gershun, A. I. Pashnev, Teor. Mat. Fiz.,
		73, 294, (1987).
\end{thebibliography}
\end{document}